\documentclass[12pt]{JHEP3}
\usepackage[dvips]{epsfig}
\usepackage{epsfig}
\usepackage{graphicx}

\def\cee{{\relax\hbox{$\inbar\kern-.3em{\rm C}$}}}

\def\third{\textstyle{1\over3}}

\def\ie{{\it i.e.,}}

\newcommand{\be}{\begin{equation}}
\newcommand{\ee}{\end{equation}}
\newcommand{\bea}{\begin{eqnarray}}
\newcommand{\eea}{\end{eqnarray}}
\newcommand{\bml}{\begin{mathletters}}
\newcommand{\eml}{\end{mathletters}}

\title{Observing braneworld black holes}
\author{Ruth Gregory$^1$, Richard Whisker$^1$ 
Kris Beckwith$^2$, and Chris Done$^2$\\
$^1$Institute for Particle Physics Phenomenology and Centre for
Particle Theory\\
$^2$Institute for Computational Cosmology\\
\\Department of Physics, University of Durham,
South Road, Durham, DH1 3LE, U.K.}

\abstract{
Spacetime in the vicinity of an event horizon can be probed
using observations which explore the dynamics of the accretion
disc. Many high energy theories of gravity lead to modifications
of the near horizon regime, potentially providing a testing ground
for these theories. In this paper, we explore the impact of 
braneworld gravity on this region by formulating a method of 
deriving the general behaviour of the as yet
unknown braneworld black hole solution. We use simple bounds to
constrain the solution close to the horizon.}

\keywords{Black holes, Extra dimensions, Braneworlds.}
\preprint{IPPP/04/33\\DCPT/04/66}

\begin{document}

\section{Introduction}

Black holes are a fascinating topic of study, whether it be
to explore the geometry of strong gravity and its quantum 
effects, or to understand the astrophysics of massive objects in
our galaxy. From the point of view of standard physics in four 
dimensions, black holes are well described by the Kerr-Newman 
family of solutions, solutions in standard four-dimensional 
Einstein-Maxwell gravity. However, the true nature of gravity, 
though well measured from tabletop to solar system scale experiments,
is surprisingly less well determined at larger or smaller scales.
For example, the assumptions of dark matter and/or energy are made
in order to fit the observed Universe with Einstein theory, yet it
is quite possible that it is the theory of gravity which should be
altered to fit these data instead.  

Theoretically, the possibility that 
gravity might not be fundamentally four-dimensional,
or indeed Einsteinian, is gathering credence. This is in
part an impact of superstring theory, which is consistent
in ten dimensions (or M-theory in eleven dimensions), but also
the more phenomenological recent developments
of ``braneworld'' scenarios, \cite{ADD,RS}, have had a direct influence on
studies of more exotic gravitational theories in four-dimensions.
Experimentally, the small scale implications of alternative
theories of gravity have focused on the impact of Kaluza-Klein (KK)
modes or small black holes in colliders, \cite{KKC,BHC}, 
cosmic ray showers, \cite{CRS}, or in particle
interactions in supernovae or nucleosynthesis, \cite{SNN}. 
Large scale implications of modified gravity have been conducted 
mostly within the MOND (modified Newtonian gravity) set-up \cite{MNG},
although some preliminary studies have investigated braneworld
modifications to the microwave background \cite{BCMB}. Interestingly, while
much theoretical work has been done on black holes in modified
or higher dimensional gravity theories, there is very little direct link
with experimental data so far.
Yet there are observations of astrophysical black holes 
which provide some constraints on strong gravity. Stellar remnant 
black holes, with masses of about 10 times the mass of our Sun, 
can form as a result of very massive
star evolution. Most stars form in binary systems, so if the companion
star is close enough then it can provide a source of material falling onto
the black hole. This material has the angular momentum of the orbit so it
forms a ring, but angular momentum transport (via a magnetic dynamo
process, \cite{BHaw}) spreads this ring out into a disc. The
magnetic dynamo also dissipates energy, so this means that there is
emitting material down to the last stable orbit around the black hole.

For large mass accretion rates this disc emission takes a rather simple
form as the material is optically thick. The energy is emitted and
absorbed many times before it escapes the disc, so thermalizes to a
blackbody spectrum at a given radius. The total disc spectrum is a sum
over all radii of these different temperature blackbodies, with the
maximum temperature (which is of order 1~keV for these stellar mass black
holes) being set by the emission from close to the inner edge of the disc
\cite{ShSun}. The combination of observed disc luminosity and
temperature can constrain the size of the inner disc radius, while
dynamical studies of the orbit can constrain the black hole mass. Together
these give an estimate of the innermost stable orbit in terms of
Schwarzschild radii. Current observations indicate that this is generally
of order $6GM$, as expected for a Schwarzschild black hole, 
though there are
a few objects with significantly smaller radii, which are interpreted as
moderate spin Kerr black holes \cite{PossK}.
                                                                                
Another observational tracer of the gravitational field comes from a
corona above the disc, which forms a high energy tail to the disc
emission. These high
energy X-rays illuminate the disc and form an iron fluorescence line,
produced from material with a large rotational velocity in a
strong gravitational field. Thus what is seen at infinity is affected by
both special (doppler shift, length contraction, time dilation) and
general (gravitational redshift) relativistic effects. Together these
transform an initially narrow atomic transition into a broad, skewed and
reddened line profile. Current observations indicate that the observed
lines are consistent with material close to the last stable
orbit in Schwarzschild or Kerr black holes \cite{FIRY}. These
relativistic effects also distort the spectrum of the intrinsic (sum of
blackbodies) disc emission \cite{Cunningham}.

Thus black holes provide a promising test bed for testing our
understanding of gravity. While stellar, or galactic, black
holes do not provide regions of high curvature, at least at the 
classical level, if we accept as fact that black holes radiate, 
then it is likely that close to the event horizon quantum effects 
become relevant. Alternatively, the effect of higher dimensions or
additional fields in the gravity sector might begin to make their
presence felt as the event horizon is approached. In this paper 
we explore the case of braneworld black holes.

The braneworld paradigm views our
universe as a slice of some higher dimensional spacetime.
Unlike the Kaluza-Klein picture of extra dimensions, where 
we do not notice the extra dimensions because they are so small
and our physics is `averaged' over them, the braneworld 
picture can have large, even non-compact, extra dimensions
which are unobservable at low energies since we are confined
to the brane. Naturally, at not so low energies, the extra
dimensions can have experimental consequences \cite{ADD,RS}.
Confinement to the brane, while at first sounding counter-intuitive,
is in fact a common occurrence. The first braneworld scenarios
\cite{EBW} used topological defects to model the braneworld, and
zero-modes on the defects to produce confinement. Of course in 
string theory, D-branes have `confined' gauge theories on their
worldvolumes.

The braneworld scenario provides a set-up in which we have standard
four-dimensional physics confined to the brane, but gravity (and possibly
a small number of other fields) propagating in the bulk. The new
phenomenology of braneworld scenarios is then primarily located
in the gravitational sector, with a particularly nice possible
resolution of the hierarchy problem being its primary motivation.
Clearly however, the astrophysical and cosmological implications 
of such a scenario have a more immediate, and more directly
measurable, impact. For these issues, the most popular model to
explore, and the one which we will be using, has been the 
Randall-Sundrum scenario, \cite{RS}, which
consists of a domain wall universe living in five-dimensional
anti-de Sitter (adS) spacetime. This model is loosely motivated
by the Horava-Witten compactification of M-theory \cite{HW}.

The Randall-Sundrum model has one (or two) domain walls situated
as minimal submanifolds in adS spacetime. In its canonical form, 
the metric of the braneworld is
\be
ds^2 = e^{-2k|z|} \left [ dt^2 - d{\bf x}^2 \right ] - dz^2
\label{rsmet}
\ee
Here, the spacetime is constructed so that there are four-dimensional
flat slices stacked along the fifth $z$-dimension, which have a 
$z$-dependent conformal pre-factor known as the warp factor. Since
this warp factor has a cusp at $z=0$, this indicates the presence of a
domain wall -- the braneworld which represents an exactly flat 
Minkowski universe.

Randall and Sundrum showed in their original papers that although 
gravity was inherently five-dimensional, and the spacetime was
strongly warped, as far as a four-dimensional braneworld observer was
concerned, the Newtonian potential of a particle on the brane was
indeed the four-dimensional $1/r$ potential. These early results 
were backed up by more complete analyses confirming that the 
graviton propagator did indeed have the correct tensor structure, 
and that the effect of the KK modes was to introduce a $1/r^3$
correction to the gravitational potential \cite{GT}.

Of course, of real interest in astrophysics and cosmology is
not the perturbative study of the graviton propagator, but a
concrete understanding of the true nonperturbative nature 
of gravity on the brane. This is what is relevant for a black hole.
To some extent, this was provided
by the brane-based approach of Shiromizu et.\ al.\ (SMS) \cite{SMS},
who used a Gauss-Codazzi approach to obtain the braneworld
gravitational equations:
\be
G_{\mu\nu} = \Lambda_4 g_{\mu\nu} + 8\pi G_N T_{\mu\nu} 
+ \kappa^2 S_{\mu\nu} + {\cal E}_{\mu\nu}
\label{smseq}
\ee
Here, $\Lambda_4$ is a residual cosmological constant on the brane. It
represents the mismatch between the brane tension and the bulk
negative cosmological constant. (In the RS scenario, the brane
tension was precisely tuned to balance out the negative bulk
cosmological constant.) The final two terms are the brane gravity
effects. The first, $S_{\mu\nu}$, consists of squares of the
energy-momentum tensor, which is only relevant at high energies.
The final term, ${\cal E}_{\mu\nu}$, is a so-called Weyl term, 
and consists of the projection of the bulk Weyl tensor onto
the brane. Although the SMS equations do provide an apparent
nonperturbative description of gravity on the brane, it is 
important to emphasize that the Weyl term is not given in terms
of data on the brane, and is a complete unknown from the brane
point of view. Therefore, in order to take a pure
brane-based approach to gravity, some assumptions must
be made about ${\cal E}_{\mu\nu}$. 

Two main nonperturbative gravitational problems are of clear interest:
Cosmology, and Black Holes. The first problem, that of finding
the braneworld generalization of the FRW universe, has been 
well explored and understood. Brane-based work concentrates 
mostly on an explanation of the ``non-conventional'' energy
momentum squared terms, as well as the unknown dark radiation
effect from the Weyl term \cite{NCC}. For cosmology however, the high
degree of symmetry present renders the full five-dimensional
problem fully integrable \cite{BCG}, and the general cosmological
braneworld is fully understood in terms of a slice of a 
five-dimensional adS black hole \cite{BCOS}. The mass of this 
bulk black hole then generates the radiation-style Weyl term. 
It is therefore
fair to say that while all the implications may not have been
calculated, brane cosmology for the pure RS scenario is pretty
well understood.

The situation for the black hole on the brane is somewhat different
however. Though it would appear that the two cases are similar in
that cosmological branes have a dependence on time as well as
the bulk $z$-coordinate, and black holes on radius and bulk
$z$-coordinate, in fact, the symmetry groups of the spacetimes
are crucially different. For cosmology, the metric splits into two
parts -- the two dimensions on which it depends, and the 
spatial part of the universe, which has constant curvature.
The mathematics of the cosmological braneworld is therefore 
a two dimensional field theory which turns out to be totally
integrable. For the black hole however, the metric splits into 
three parts -- the two dimensions on which it depends, the time 
coordinate and the remaining spatial part in which the horizon 
resides. Thus there are two fields in our two dimensional theory, 
one of which acquires a potential, and there is
no longer a simple solution \cite{CG}.

The case of the black hole in this braneworld picture now
becomes of great interest and importance. Of interest as a problem
in higher dimensional gravity, and of importance because the
adS/CFT correspondence, \cite{MAL}, relates the classical five-dimensional
braneworld black hole solution to the four-dimensional
quantum radiating black hole\cite{DL,EFK}.  
The first attempt, \cite{CHR}, 
to find a black hole solution replaced the Minkowski metric 
in (\ref{rsmet}) by the Schwarzschild metric, thus creating a 
black string sticking out of the brane.  Unfortunately, as 
suspected by the authors, this string is unstable to classical 
linear perturbations \cite{BSINS}. Chamblin et.\ al.\ realised
that the true localised black hole would be a slice of a 
five-dimensional accelerating black hole metric (known as 
the C-metric, \cite{CMET}, in four dimensions), however no
such metric has as yet been found. A lower dimensional
version of a black hole living on a $2+1$-dimensional braneworld
was however presented by Emparan, Horowitz and Myers \cite{EHM},
using this four-dimensional C-metric. Since then, several authors
have attempted to find the full metric -- notably numerical work
by Wiseman \cite{TW}, and others \cite{PAE}.

In this paper, we are interested in near horizon modifications to
General Relativity, which in the absence of a full five-dimensional
solution might seem problematic. However, our aim here is not to
attempt to answer the full five-dimensional problem, but to revisit 
the four-dimensional braneworld and to take a practical 
approach to finding the braneworld metric. The motivation will be
to see if we can categorize (preferably analytically)
various classes of near horizon behaviour,
and to see if there is any universality to the four-dimensional
braneworld solutions.

The literature has several special solutions, notably 
the tidal Reissner-Nordstrom solution of Dadhich et.\ al.\ \cite{DMPR}, 
but also the PPN parametrized solutions of Casadio et.\ al.\ \cite{CAS}. 
Visser and Wiltshire \cite{VW} presented a more
general method which generated an exact solution
for a given radial metric form. All of these approaches however 
have in common the assumption of an ``area gauge'' for the 
radial coordinate, $r$, in the metric, \ie\ that the area 
of the 2-spheres surrounding the black hole behaves 
as ${\cal A}(r) = 4\pi r^2$. Thus the area of spheres surrounding 
the black hole increases monotonically between the horizon and 
spacelike infinity. However, the monotonicity of ${\cal A}(r)$ 
is only guaranteed if the dominant energy condition holds, and 
there is no reason to suppose that this will be the case for the 
Weyl term, indeed, Dadhich et.\  al.\  have ${\cal E}_{00}<0$ -- a 
violation of the weak energy condition! In fact, energy 
conditions are persistently violated in dimensionally reduced 
theories of gravity.

Our reason for suspecting that ${\cal A}(r)$ is not monotonic 
lies in the putative higher dimensional C-metric, which would
consist of an accelerating black hole being `pulled' by a string.
The appropriate higher dimensional metric for a Poincar\'e invariant
string has a turning point in the area function, and the `horizon' 
is in fact singular \cite{CPB}. It is therefore possible that
this renders the black hole horizon also singular. Moreover,
using the dual description of a CFT with cutoff living on the 
brane \cite{DL}, a static quantum black hole must have a 
singular horizon \cite{EFK}. Since the
singularity of the string has a diverging area function, it seems
likely that the black hole itself might. Such a turning point in 
the area function can be thought of as a {\it wormhole} in the
geometry, so called because the spatial part of the Schwarzschild 
metric:
\be
|ds^2| = \left ( 1-{2GM\over r} \right )^{-1} dr^2 + 
r^2 d\Omega^2_{I\!I} = {({\tilde r}+GM/2)^4\over{\tilde r}^4}
d{\tilde r}^2 + {({\tilde r}+ GM/2)^4 \over
{\tilde r}^2} d\Omega^2_{I\!I}
\ee
(where $r = ({\tilde r} + GM/2)^2 /{\tilde r}$)
has this property of the area of spheres decreasing as we approach
$r=2GM$, then stationary at $r=2GM$, then increasing again as we
move onto the other asymptotically flat r\'egime of the maximally
extended Schwarzschild solution. As we will see, this is a very good 
analogy, as there is an exact analytic braneworld metric which
has this spatial form, and simply moves the event horizon relative
to Schwarzschild either outside $r=2GM$ or through the wormhole neck
onto the other Kruskal branch.

In identifying general features of braneworld solutions,
the questions we explore in this paper are:

$\bullet$ When is the horizon singular?

$\bullet$ When does the area function have a turning point?

$\bullet$ When are black hole solutions asymptotically flat?

Within the context of an equation of state of the Weyl
term we are able to give definitive answers to all of these
questions.

The layout of the paper is as follows. In the next section we
derive the braneworld equations in a general gauge and show
how to reduce these to a two-dimensional dynamical system for
a given equation of state. At this point we give the analytic solution
corresponding to the Schwarzschild wormhole.
In section \ref{sec:gends} we analyze this system in
general, showing how to answer the questions above. 
In section \ref{sec:physsol} we make contact with the asymptotic 
linearized propagator, and work on small black holes, and present
arguments leading to an analytic near horizon metric. 
Finally, in section \ref{sec:bhsol}, we use this metric to 
explore the phenomenology of astrophysical black holes.

\section{Spherically symmetric braneworld metrics}\label{sec:genform}

We start by looking at the general static, spherically symmetric metric
on the braneworld. By taking the metric to be static, we are taking
the point of view that there exists a five-dimensional solution
analogous to the C-metric in four-dimensions which has a timelike 
Killing vector, and can therefore be `sliced' by the braneworld in
such a way as to create a static four-dimensional black hole on
the brane. In this we assume that the three-dimensional braneworld
black hole constructed from the four-dimensional C-metric has a
direct analog in one dimension higher. We note that this produces
a static braneworld black hole, and precludes the possibility 
of a dynamical back-reaction to Hawking radiation,
instead corresponding to a Boulware choice of boundary conditions,
\cite{EFK}, which give a singular `horizon'.
It should be noted that there is not a consensus as to whether the
braneworld black hole should be static (and therefore singular, by the
reasoning of \cite{EFK}), or time dependent, corresponding to an
evaporating black hole, as explored by Tanaka \cite{TAN}.

\subsection{The metric}
The general static spherically symmetric
metric on the brane can be written as:
\be
ds^2 = A^2(r) dt^2 - B^2(r) dr^2 
- C^2(r) d\Omega_{I\!I}^2 
\label{genmet}
\ee
Clearly this is not in the simplest gauge, as we can still choose 
our radial function, $r$, quite arbitrarily, however, it proves to 
be convenient to use this over-general form in order to choose the
best gauge for problem solving, and to compare this to more familiar
gauges more readily. The main reason for using this form of the metric,
with an arbitrary function for the area of the 2-spheres rather than
$r^2$, is that there is good reason to believe that the area function 
might not be monotonic. With this form of the metric, the second 
derivative of the area radius, $C$, (\ie\ the radial function defined
by $\sqrt{{\cal A}/4\pi}$) is given by the following combination of the 
Einstein tensor:
\be
{C''\over C} = -{B^2\over 2} \left [ G^t_t - G^r_r \right]
+ {C'\over C} \left ( {B'\over B} + {A'\over A} \right )
\ee
therefore, for the area function to be guaranteed to be monotonic,
we must have $ G^t_t - G^r_r \geq 0$, which is equivalent to the
dominant energy condition. In usual Einstein gravity this is generally
the case, but when there are extra dimensions, or extra
fields, this is no longer the case. For example, in the cosmic $p$-brane,
\cite{CPB} the area function actually blows up on the `horizon'. Since
the cosmic $p$-brane might be expected to have some connection to the 
higher dimensional C-metric, possibly rendering the `horizon' singular,
it is vital that in any exploration of braneworld black holes we do not
make the restrictive ansatz of $C=r$. 

The `vacuum' brane equations from (\ref{smseq}) are
\be
G_{\mu\nu} = {\cal E}_{\mu\nu}
\ee
We follow Maartens \cite{MM} in using the symmetry of the physical
set-up to put the Weyl energy into the form:
\be
{\cal E}_{\mu\nu} = {\cal U} \left ( u_\mu u_\nu - {\third} h_{\mu\nu}
\right ) + \Pi \left (r_\mu r_\nu + {\third} h_{\mu\nu} \right )
\ee
where $u^\mu$ is a unit time vector, and $r^\mu$ a unit radial vector.
Note that the Weyl energy is in `Planck' units, \ie\ there is no preceding
$8\pi G$ since ${\cal E}$ is derived from gravitation in the bulk. If we 
want to compare with 4D matter, we should rescale by $1/8\pi G$.

We now have the equations of motion:
\bea
G^t_t &=& {1\over C^2} - {1\over B^2} \left [2{C''\over C} 
- 2 {C'B'\over CB} + {C^{\prime2}\over C^2} \right ]
={\cal U} \label{Eintt} \\
G^r_r &=& {1\over C^2} - {1\over B^2} \left [ 
2{A'C'\over AC} + {C^{\prime2}\over C^2} \right ] = - {({\cal U} + 2\Pi)
\over 3} \label{Einrr} \\
G^\theta_\theta  &=& -{1\over B^2} \left [ 
{A''\over A} + {C''\over C} + {A'C'\over AC}
- {A'B'\over AB} - {B'C'\over BC} \right]  
= - {({\cal U} - \Pi ) \over 3} \label{Einth} 
\eea
An alternate and useful equation is the Bianchi identity:
\be
\left ( {\cal U} + 2\Pi \right )' + 2 {A'\over A} \left ( 
\Pi + 2{\cal U}  \right ) + 6\Pi {C'\over C} =0
\label{Bid}
\ee
(or conservation of ``energy-momentum''). 

This system of equations has been solved in many special cases,
and more general techniques have also been presented.
Briefly, the special cases are the tidal Reissner-Nordstrom 
solution of Dadhich et.\ al.\ \cite{DMPR}, or the solutions which
assume a given form for the time or radial part of the metric
\cite{SSOL}. Visser and Wiltshire \cite{VW} presented a more
general method which generated an exact solution
for a given radial metric form. In all of these cases however, the 
radial gauge $C=r$ was chosen (although \cite{VW} did comment on
how to use their method when $C$ was not monotonic).

A reasonable alternative to making guesses for various of the metric
functions is instead to follow a pragmatic approach as in cosmology.
When solving for an FRW universe, the precise details of the 
composition of the universe are approximated by an isotropic perfect 
fluid energy-momentum tensor, and, most pertinently, an equation of state
is assumed for this source. Dust ($p=0$) for the later universe, 
radiation ($p=\rho/3$) for somewhat earlier times, and cosmological
constant ($p=-\rho$) for inflation. Clearly the actual evolution and
matter content of the universe is more detailed and complicated, but
this method is useful, accurate, and universally accepted. Here we 
propose the analogous approach: an equation of state for ${\cal U}$ 
and $\Pi$, $\Pi = {\gamma-1\over2}{\cal U}$. Of course, a priori there
is no reason to suppose that the Weyl term should obey an equation 
of state (and indeed we will argue later that it will not)
however, it is quite possible that just as the universe is approximately 
matter dominated or radiation dominated in certain eras, the Weyl term
might have certain asymptotic equations of state which may be useful as
near-horizon or long range approximations.  

\subsection{The dynamical system}

In the next section we provide a detailed analysis of all equations
of state, using the common method of
dynamical systems analysis of the equations of motion. In our case,
it is clear that the most convenient gauge for this type of analysis is
$B=C$. In this case, writing
\be
X = {C'\over C} + 2 {A'\over A} \qquad Y = {C'\over C}
\ee
we have:
\bea
2X'+X^2-1 &=& 3{\cal U} C^2 = {9\over\gamma}
(XY-1) \label{Xeq}
\\
1-2Y'-Y^2 &=& {\cal U} C^2 = {3\over\gamma}
(XY-1) \label{Yeq}
\eea
together with the constraint
\be
1 - XY = - {C^2\over 3} ({\cal U}+2\Pi)
= - \gamma {{\cal U} C^2 \over 3}
\label{coneq}
\ee
The plot of the $\{X,Y\}$ phase plane gives us curves which are
solutions, $X(\rho), Y(\rho)$, to this dynamical system.
Whether or not these trajectories
correspond to an actual black hole depends on whether the integrated
solutions $A(\rho), C(\rho)$ have the behaviour we expect for a black
hole, such as an horizon, asymptotic flatness and so on. 

It is therefore useful before proceeding with the analysis to extract
some asymptotic information from the Einstein equations in order to
identify general features of the solution on the phase plane.

\subsection{Asymptotic analysis}

There are two clear asymptotic regions in which we would like to 
have some information about the black hole solution - the horizon
and infinity. Near the horizon $A\to 0$, and we therefore expect
that $X$ will become infinite. Near infinity however we would like
spacetime to be asymptotically flat, which means (in the area gauge)
$A \sim 1 - O(r^{-1})$, $B\sim 1 - O(r^{-1})$, $C=r$. Or 
$A \sim 1 - O(e^{-\rho})$, $B=C\sim e^\rho +O(1)$ in dynamical
system (d.s.) gauge.

For the purpose of clarity, we will look at the far field region 
in the area gauge, in which the leading behaviour of the Weyl 
energy and radial metric component is easily read off as:
\bea
{\cal U} &=& {u_0 r^{3\over \gamma}\over r^3} \label{farfu} \\
B^{-2} &=& 1 - {\mu\over r} - {\gamma u_0 r^{3\over\gamma}\over 3r}
\label{farfB}
\eea
What this clearly shows is that for $0<\gamma<3$ the braneworld metric
does not tend to Minkowski spacetime for large $r$. Indeed, even for
$\gamma>3$ the metric is not asymptotically flat, since asymptotic 
flatness requires
corrections to Minkowski spacetime to be $O(1/r)$. Therefore, an
asymptotically flat braneworld black hole requires an asymptotic
equation of state $\gamma<0$.

Near the horizon on the other hand, we cannot assume the area gauge,
and integrating the energy conservation equation gives
\be
{\cal U} \propto A^{-{3\over\gamma}-1} C^{{3\over\gamma}-3}
\label{ubehav}
\ee
Whether or not the horizon corresponds to a singularity in the
Weyl energy depends on the magnitude and sign of $\gamma$, as well
as on the behaviour of $C$.

\subsection{A simple analytic solution}\label{sub:ansol}

We conclude this section by demonstrating the use of both the
general gauge $C(r)$, as well as the dynamical system by deriving
an alternate form of a known analytic solution with ${\cal U}=0$.
A quick glance at the equation of state shows that this is formally
the limit $\gamma\to\pm\infty$. Since letting $\gamma$ become infinite
decouples the equation of state from the dynamical system, the
phase plane is in fact the same whether $\gamma$ is $+\infty$
or $-\infty$. Our main reason for deriving this solution afresh is
that we wish to use it in section \ref{sec:bhsol} for a near horizon
limit, but clearly, any exact solution is helpful, and our derivation
makes the overall structure of the spacetime somewhat clearer than
using an area gauge.

For ${\cal U}=0$, (\ref{Xeq},\ref{Yeq}) have the solutions:
\be
X,Y = \tanh {(\rho-\rho_0)\over 2} \ , \coth{(\rho-\rho_1)\over2}
\label{XYan}
\ee
Clearly the former solution is appropriate for $Y$, as this corresponds to
\be
C = C_0 \cosh^2 {(\rho-\rho_0)\over 2}
\label{Can}
\ee
which, since we expect an asymptotically flat spacetime which has
$r=e^\rho$, gives $C_0=4e^{\rho_0}$. We then have for $A$:
\be
2 \ln A + \ln C = 2\ln \cosh {(\rho-\rho_1)\over2} 
\ee
or
\be
2 \ln A + \ln C = 2 \ln \sinh {(\rho-\rho_1)\over2}
\ee
i.e.
\be 
A^2 A_0^{-2} C_0 \cosh^2 {(\rho-\rho_0)\over 2}
= \cosh^2 {(\rho-\rho_1)\over2} \ \ {\rm or} \ \ 
\sinh^2 {(\rho-\rho_1)\over2}
\ee

For an horizon, we expect that $A^2$ will have a zero, hence we
choose the second solution, giving
\be
A^2 = {A_0^2\over C_0} \left ( {\cosh(\rho-\rho_1) - 1\over
\cosh (\rho-\rho_0) + 1} \right )
\label{Aan}
\ee
For asymptotic flatness $A_0^2 = 4 e^{\rho_1}$.

Overall the metric is:
\be
ds^2 = e^{\rho_1-\rho_0} \left ( {\cosh(\rho-\rho_1) - 1\over
\cosh (\rho-\rho_0) + 1} \right )
dt^2 -
16 e^{2\rho_0} \cosh^4 {(\rho-\rho_0)\over2} 
\left [ d\rho^2 + d \Omega_{I\!I}^2 \right ]
\ee
or, writing $r=e^\rho$:
\be
ds^2 = {(r-r_1)^2\over (r+r_0)^2} dt^2 - {(r+r_0)^4\over r^4} dr^2
- {(r+r_0)^4\over r^2} d\Omega_{I\!I}^2
\label{wormmet}
\ee
this metric has appeared in the area gauge as \cite{CAS}
\be
ds^2 = \left [ (1+\epsilon) \sqrt{1-{2GM\over R}} - \epsilon \right ]^2
dt^2 - {dR^2 \over 1-{2GM\over R}} - R^2 d\Omega_{I\!I}^2
\label{areaworm}
\ee
where $R= (r+r_0)^2/r$, $GM=2r_0$, and $GM\epsilon = r_1-r_0$.
The anisotropic stress for this solution is $\Pi= 3GM\epsilon/AC^3$.
\FIGURE{
\includegraphics[height=5cm]{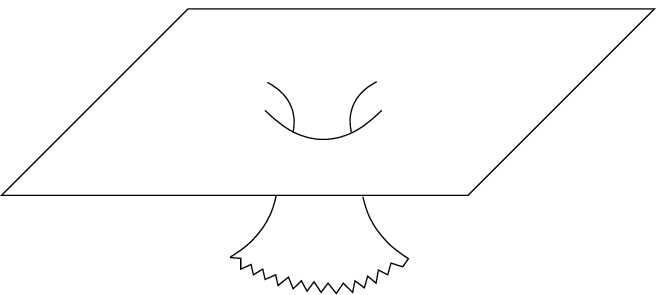}
\caption{Pictorial representation of a constant time slice of the 
metric (\ref{wormmet}).} 
\label{fig:worm}
}

While the area gauge gives a familiar spatial part of the metric, note
that at the old Schwarzschild `horizon', $R=2GM$, the $t-t$ component of
the metric does not vanish. For $\epsilon>0$, $g_{tt}$ will be zero before
$r=2GM$, and the area gauge holds outside the black hole. (Strictly of course,
this is not a black hole, as the `horizon' is singular.) If $\epsilon <0$ 
then $g_{tt}=0$ at $r<2GM$, and $r=2GM$ becomes a coordinate singularity
accessible by timelike observers,
the result of choosing an inappropriate gauge.

\section{The general dynamical system}\label{sec:gends}

In order to find a general solution for the braneworld black hole,
subject to the proviso that the Weyl energy obeys an equation of state, 
we will now analyze the general dynamical system:
\bea
X' &=& {1-X^2 \over 2} + {9\over 2\gamma} (XY-1) \label{XX} \\
Y' &=& {1-Y^2\over2} + {3\over 2\gamma} (1-XY) \label{YY}
\eea
Once again we emphasize that we do not believe that the equation of
state will necessarily hold over the whole horizon exterior, simply 
that it will potentially give various asymptotic behaviours near infinity
or near the horizon.

The plot of the $\{X,Y\}$ phase plane gives us curves which are
solutions, $X(\rho), Y(\rho)$, to the dynamical system (\ref{XX},\ref{YY}).
These in turn can be integrated to find $A(\rho), C(\rho)$. A selection of
representative phase plane plots are given in figure 
\ref{fig:inf}. 
\FIGURE{
\includegraphics[height=15cm]{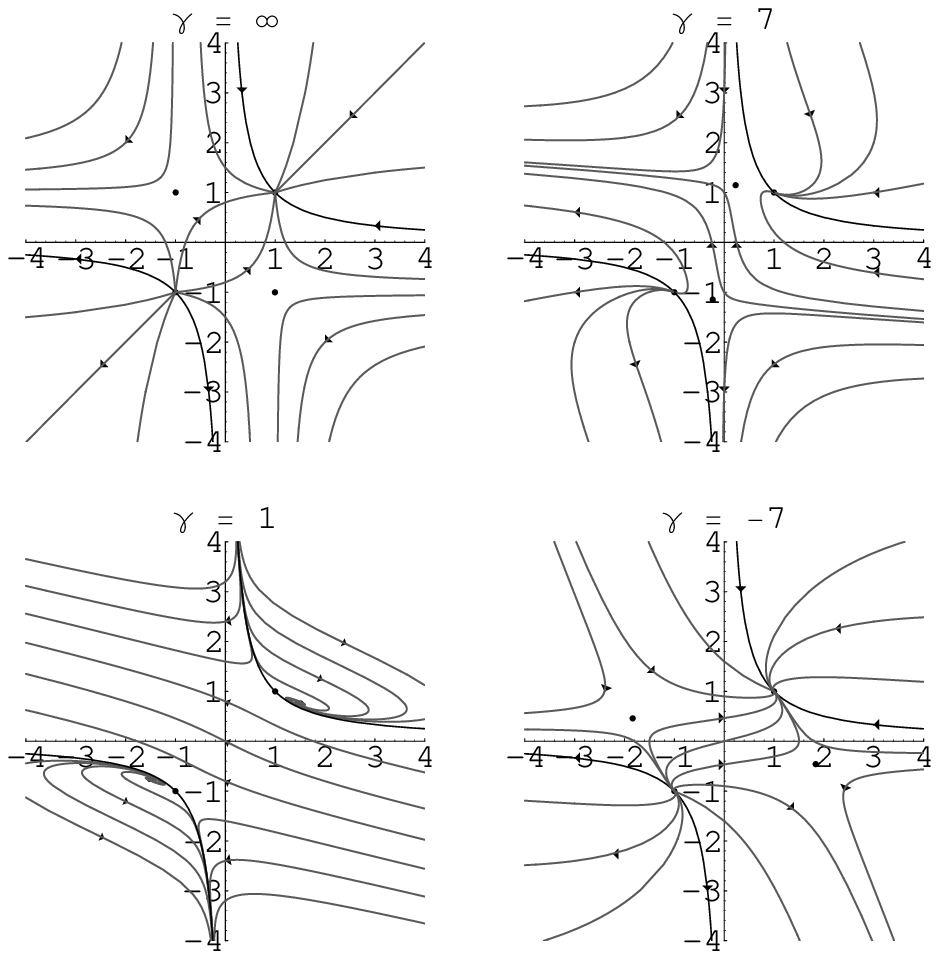}
\caption{Phase plane for a range of equations of state.}
\label{fig:inf}
}
What these plots show is that there are
attractors in the phase plane, and that a general trajectory flows in from
infinity to one of these attractors. 
Whether or not these trajectories
correspond to an actual black hole depends on whether the integrated
solutions $A(\rho), C(\rho)$ have the behaviour we expect for a black
hole, such as an horizon, asymptotic flatness and so on. 

Briefly, a horizon corresponds to $A\to0$,
therefore we expect $X$ becomes infinite for the black hole horizon.
Flat space on the other hand corresponds to $A'=0$, $C \propto e^\rho$, 
hence $X=Y=1$. 
A black hole solution must therefore be a trajectory which comes in from
large $X$ (possibly large $Y$ as well) and terminates on $(1,1)$. To
find out which equations of state allow this, and whether the area function
has any turning points, which correspond to zeros of $Y$, requires a
detailed analysis of the phase plane.

\subsection{Features of the phase plane}

Part of any dynamical systems analysis is an identification and 
classification of critical points, invariant submanifolds, and
other distinguishing features of the phase plane. 
In this case, the invariant hyperboloid is easily
identified from the constraint (\ref{coneq}) as $XY=1$. Along this
hyperboloid $({\cal U} + 2\Pi) = 0$, therefore, apart from the
special case $\gamma=0$, this corresponds to a vanishing Weyl
term, and hence the pure Einstein case -- \ie\ the Schwarzschild
solution. This is given by (\ref{Can},\ref{Aan}) with $\rho_0=\rho_1$
and $A_0^2=C_0$. If $\gamma=0$, in addition to the Schwarzschild solution
we have $X=\pm\sqrt{3}=1/Y$, which is covered by the analysis of the
critical points below.

\vskip 2mm
                                                                                
\noindent{\it Critical points}
                                                                                
\vskip 1mm

The system has 4 critical points ($X'=Y'=0$):
\bea
P_\pm &=& \pm (1,1)\\
Q_\pm &=& \pm (27 + \gamma^2) ^{-1/2} (9 - \gamma, 3 + \gamma )
\eea
The $Q_\pm$ critical points move as $\gamma$ runs from $+\infty$ to
$-\infty$, from $Q_\pm=(\mp1,\pm1)$ to $Q_\pm=(\pm1,\mp1)$. 
For $\gamma=3$, $P$ and $Q$
are coincident. The nature of the critical points is as follows:

\vskip 1mm

\noindent $\bullet$ $3<\gamma$ \hskip 11mm $P_+$ is an attractor 
and $P_-$ a repellor, the $Q$'s are saddle points. 

\vskip 1mm

\noindent $\bullet$ $0<\gamma<3$ \hskip 3mm $Q_+$ is an attractor 
and $Q_-$ a repellor, the $P$'s are saddle points.

\vskip 1mm

\noindent $\bullet$ $\gamma<0$ \hskip 11mm $P_+$ is an attractor 
and $P_-$ a repellor, the $Q$'s are saddle points. 

\vskip 1mm

The critical point $(1,1)$ corresponds to flat space:
\bea
X=Y=1 &\Rightarrow& A'=0,\ C'=C \nonumber \\
&\Rightarrow& C=e^\rho
\eea
Therefore any asymptotically flat solution must terminate on this
critical point. While $P_+$ is an attractor, this is not really a problem,
but for the range of $\gamma$ where it is a saddle point, only the 
invariant hyperboloid can satisfy this, and by definition, this is
where we have the exact Schwarzschild solution.

The critical point $Q_+$ on the other hand, corresponds to a 
non-asymptotically flat spacetime, which, for $\gamma>-3$ in
area gauge is:
\be
ds^2 = r^{2(3-\gamma)\over(3+\gamma)} dt^2 
- {(27+\gamma^2)\over(3+\gamma)^2} dr^2 - r^2 d\Omega_{I\!I}^2
\ee
which we do not expect to be appropriate to the metric for an 
isolated source. This solution can also be used for $\gamma<-3$,
provided one remembers that increasing $r$ actually corresponds
to moving {\it towards} the black hole ($g_{tt}\to 0$ as
$r\to\infty$). This is a genuine wormhole solution, in that the area
${\cal A}(r)$ increases unboundedly towards the event horizon,
which is located at infinite proper distance. Unlike the
Schwarzschild wormhole however, this inner asymptotic region is
not flat, but, as already mentioned, leads in to a null asymptopia.
It is perhaps worth noting that the critical value $\gamma=-3$
corresponds to the marginal case of no wormhole, but an infinite
throat:
\be
ds^2 = r^{2} dt^2 
- {dr^2\over r^2} - r_0^2 d\Omega_{I\!I}^2
\ee
exactly analogous to the extreme Reissner-Nordstrom black hole throat.

The general solution of the braneworld black hole therefore requires 
$\gamma>3$ or $\gamma<0$ in order to terminate on the critical point
$P_+$. Luckily, this range of $\gamma$ is precisely that for which
spacetime asymptotes flat space, although only for $\gamma<0$ is it
actually asymptotically flat. 

\vskip 2mm
                                                                                
\noindent{\it Asymptotes}
                                                                                
\vskip 1mm
                                                                                
Finally, the other region of interest, the black hole horizon, 
corresponds in general to large values of $X$, for which we can
identify the characteristic behaviour.

The line $Y=(\gamma-3)X/(\gamma+9)$ is an asymptote for 
$\gamma\in[0,3]$, but a separatrix for negative $\gamma$, 
and $\gamma>3$. It corresponds to the solution
\be
ds^2 = (\rho-\rho_0)^{24\gamma\over\gamma^2+27} dt^2 -
(\rho-\rho_0)^{4\gamma(\gamma-3)\over \gamma^2+27} \left [
d\rho^2+d\Omega^2_{I\!I} \right]
\label{hormet03}
\ee
which has a singular horizon.

For negative $\gamma$, and $\gamma>3$ the asymptotic solution is
$X=2/(\rho-\rho_0)$, and $Y\propto (\rho-\rho_0)^{-3/\gamma}$.
This gives the metric
\be
ds^2 = (\rho-\rho_0)^2 dt^2 - C_0^2 \left(1+c_1
(\rho-\rho_0)^{1-3/\gamma} \right )\left [
d\rho^2+d\Omega^2_{I\!I} \right]
\label{hormet3n}
\ee
The horizon is singular in this case for $|\gamma|>3$.

\subsection{Special solutions}

Because of the number and nature of the critical points, there can be
special solutions which start on one critical point and terminate
on another. Depending on the value of $\gamma$, the attractor $P_+$ 
can have up to three special solutions corresponding to trajectories 
from each of the other critical points.
It is easiest to demonstrate these solutions for the
extreme case $\gamma=\infty$, where we have an analytic solution
of the phase plane.

\vskip 2mm

\noindent$\bullet$ $P_-$ solution:

\vskip 1mm

The solution from $P_-$ to $P_+$ is easy to define in the analytic
case: it is $X=\tanh(\rho-\rho_1)/2$, $Y=\tanh(\rho-\rho_0)/2$,
which corresponds to a two parameter family solution
of metrics:
\be
\label{worm}
ds^2 = {(r+r_1)^2\over(r+r_0)^2} dt^2 
- {(r+r_0)^4\over r^4} dr^2 - {(r+r_0)^4
\over r^2} d\Omega^2_{I\!I}
\ee
For $r_0=r_1$, this is the Schwarzschild wormhole, and corresponds to the
straight line path between $P_-$ and $P_+$. For $r_1\neq r_0$, the solution
corresponds to the other paths shown in figure \ref{fig:inf} between
the critical points. The wormhole is supported
here by the negative anisotropic stress $\Pi = -6/C^3=
-6r^3/(r+r_0)^6$. See \cite{Bronn} for more detailed discussions of 
wormhole solutions
on braneworlds. Although for finite $\gamma$ the form of this
metric will change, the general nature -- that of a wormhole
connecting two asymptotically flat regions -- will not. This
solution exists for $\gamma>9$ and $\gamma<-3$.

\vskip 2mm

\noindent$\bullet$ $Q_-$ solution:

\vskip 1mm

This solution only exists in the $\gamma=\infty$ limit. It 
has $X\equiv 1$ and $Y=\tanh\rho/2$ which integrates to the metric
\be
ds^2 = {r^2\over(r+r_0)^2} dt^2 - {(r+r_0)^4 \over r^4} dr^2
- {(r+r_0)^4\over r^2} d\Omega^2_{I\!I} \label{Qmsoln}
\ee
which can be seen to be a limiting form of the metric (\ref{wormmet})
\ie\ a wormhole with an inner asymptotic null asymptopia.
For $\gamma<-3$, the $Q_+$ critical point has $Y<0$ and hence the
$Q_+\to P_+$ solution actually takes this qualitative form.

\vskip 2mm

\noindent$\bullet$ $Q_+$ solution:

\vskip 1mm

This solution has $Y\equiv 1$ and $X=\tanh\rho/2$ for the analytic
form. This integrates to the metric
\be
ds^2 = \left ( 1+ {1\over r} \right)^2 dt^2 - dr^2
- r^2 d\Omega^2_{I\!I}
\ee
which is not a wormhole, but a flat 3-space with a distorted
Newtonian potential. The solution will have this form for $\gamma>3$,
but for $\gamma<-3$, the $Q_+$ critical point moves below the 
$X$-axis, and the solution joining $Q_+$ to $P_+$ takes the above
form (\ref{Qmsoln}). For $\gamma=-3$ this trajectory is identical
to the extremal Reissner-Nordstrom solution:
\be
ds^2 = \left ( 1-{r_0\over r}\right )^2 dt^2 - 
\left ( 1-{r_0\over r} \right )^{-2} dr^2 - r^2 d\Omega^2_{I\!I}
\ee

In addition, there are special values of $\gamma$ for which there are
$Q_- \to Q_+$ trajectories. Specifically, $\gamma=9$ has a wormhole:
\be
ds^2 = {r\over (r+r_0)^2} dt^2 - {3\over4}\left (
{r+r_0\over r} \right )^4 dr^2 - {(r+r_0)^4\over r^2} d\Omega^2_{I\!I}
\ee
and $\gamma=-3$ has a `throat' solution with bouncing
Newtonian potential:
\be
ds^2 = A_0^2 \cosh^2 r\ dt^2 - dr^2 - R_0^2 d\Omega^2_{I\!I}
\ee
or adS$_2\times S^2$.

\subsection{Summary}

To summarize: the phase plane shows clearly that flat spacetime
is a critical point, which is an attractor for equations of state
with $\gamma>3$ or $\gamma<0$. Only for this latter range of 
$\gamma$ is the spacetime asymptotically flat. For these
ranges of $\gamma$ the near horizon behaviour is also given by
a defined class of metrics (\ref{hormet3n}), which are singular
for $|\gamma|>3$. Solutions are allowed both with and without turning
points in the area function, though only for $\gamma<-3$ do these
correspond to solutions which are asymptotically flat. 

In addition, there are special solutions which correspond
to trajectories between critical points. The trajectory from $P_-$ 
to $P_+$ represents a 
solution with two asymptotic flat regions, and is a genuine
braneworld wormhole. The solution from one $Q$ critical point 
also represents a wormhole spacetime, which in the case of $\gamma=-3$, 
has a throat.

\section{Physical solutions}\label{sec:physsol}

In order to decide what a reasonable black hole metric might look like, 
it is useful to compare to known results, that of the linearized
metric, and also the small black hole approximation.

\subsection{Comparison with linearized propagator}

One region in which we do know the metric of the black hole is
at large $r$. Randall and Sundrum, \cite{RS}, computed the corrections
to the Newtonian potential, and Garriga and Tanaka, \cite{GT},
the correct tensor propagator:
\be
g_{tt} = 1 - {2GM\over {\tilde r}} \left ( 1 
+ {2\over3}{l^2\over {\tilde r}^2} \right ) \hskip 5mm ;
\hskip 8mm
g_{ij} = -1 - {2GM\over {\tilde r}} \left ( 1 
+ {1\over3}{l^2\over {\tilde r}^2} \right )
\ee
where we write ${\tilde r}$ to distinguish from the $r$-coordinate of
the area gauge. Transforming to area gauge gives:
\be
g_{tt} = A^2 = 1 - {2GM\over r} - {4\over3} {GMl^2\over r^3} \hskip 5mm
; \hskip 8mm
-g_{rr}^{-1} = B^{-2} = 1 - {2GM\over r} - {2GMl^2\over r^3} 
\ee
Substituting into the equations of motion gives an equation of
state $\Pi = 5{\cal U}/4$, or $\gamma = -3/2$.

Interestingly, this asymptotically flat solution holds down to 
the event horizon, giving a fully nonsingular solution. The Weyl energy:
\be
{\cal U} = {-4GMl^2\over r^5} \left ( 1 - {2GM\over r} 
- {4\over3} {GMl^2\over r^3} \right )
\ee
has a maximum at $r\simeq 2.4GM$ and returns to zero at the event
horizon. This is not the behaviour we would expect from the Weyl 
energy, hence $\gamma= -3/2$ should be taken only as an asymptotic
equation of state.

\subsection{Comparison with small black holes}

Another common limit used in braneworld black holes is the small
black hole limit. Here, if the black hole has mass much less than
the adS scale, it is expected that the adS curvature has very little
effect on the event horizon, and so the 
five-dimensional Schwarzschild solution in a good approximation, 
i.e.\ $B^{-2} = A^2 = 1 - {\mu\over r^2}$ in area gauge. A quick 
check of the asymptotic solution (\ref{farfB}) shows that $\gamma=-3$,
or $\Pi =-2{\cal U}$. This has ${\cal U} = -\mu/r^4$.

A plot of the phase plane for $\gamma=-3$ shows that these 
\FIGURE{
\includegraphics[height=10cm]{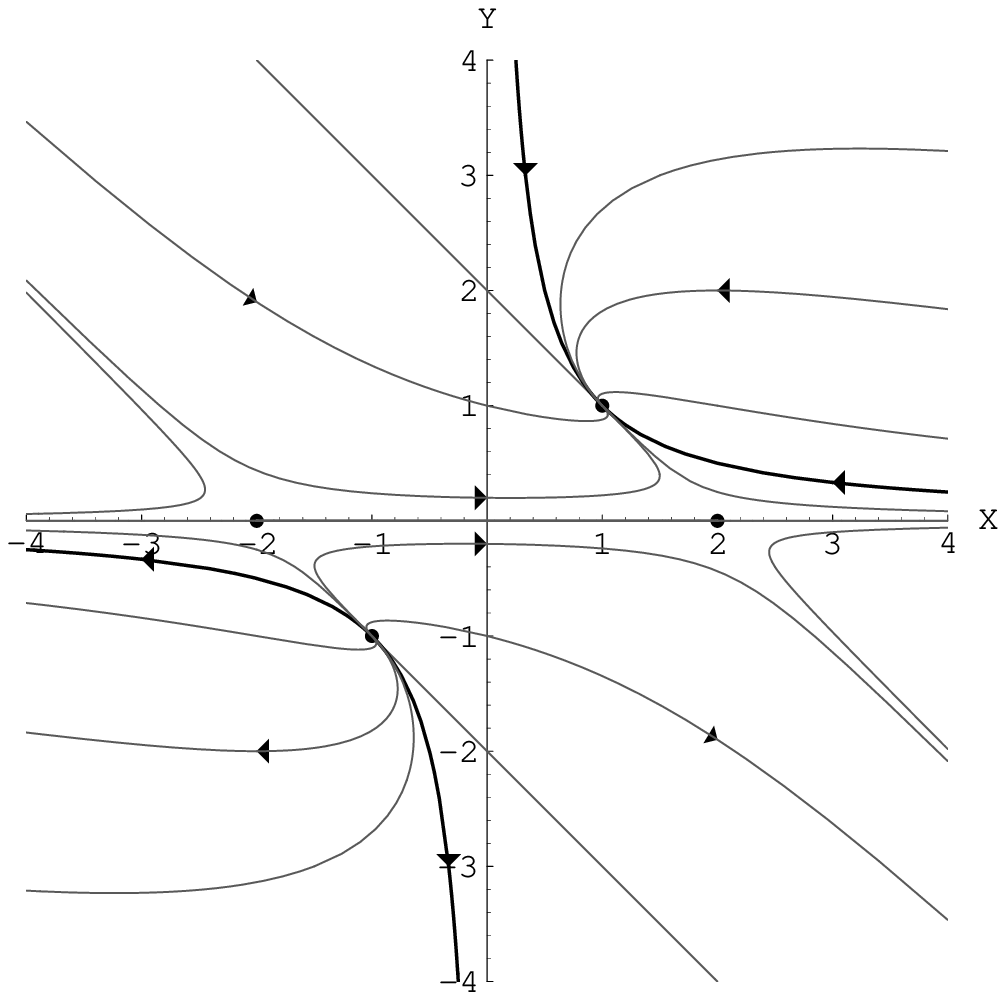}
\caption{Phase plane for $\gamma=-3$.} 
\label{fig:mthree}
}
trajectories form a stable family of solutions terminating on
$P_+$. There are no trajectories crossing the $X$-axis as $Y=0$ 
is a solution to the equations of motion. In other words, there are
solutions to the black hole system which represent truly infinite 
throats with varying Newtonian potentials. For $\gamma<-3$ once again
there are trajectories which cross the $Y$-axis, and hence have 
a bounce in the area function, or a wormhole. 

\subsection{A possible black hole solution}

Having compared the equation of state method with known asymptotic
solutions, it is clear that if an equation of state applies, it is 
generally a negative one. Moreover, from the intuition gained by
looking at very small black holes, $\gamma$ curiously becomes more
negative closer to the horizon. What sort of equation of state might
hold near the event horizon? To explore this, we return to the 
holographic correspondence.

In string theory, it has been realized for some time that there
is a correspondence between string theory on adS space, and a
CFT on the boundary of that adS space \cite{MAL}. In other 
words, all of the information contained in the five-dimensional 
gravitational spacetime is encoded in a pure quantum field theory 
(no gravity) living on a four-dimensional spacetime. 
In the braneworld picture, the brane is not at the adS boundary, 
but at a finite distance, and the theory on this brane now 
contains gravity, as well as a conformal energy-momentum tensor 
-- the Weyl term ${\cal E}_{\mu\nu}$. The effect of the brane 
on the adS/CFT correspondence therefore is that the bulk theory 
of gravity in five dimensions corresponds to the four dimensional 
brane theory of a CFT with a UV cutoff interacting with gravity 
\cite{DL}.  Since the brane theory is a quantum 
theory, the holographic correspondence indicates that the classical
bulk solution projects to a quantum corrected solution on the
brane \cite{EFK}. 

For cosmological solutions, there is a nice holographic interpretation,
where the braneworld cosmology can have a radiation source which is 
the result of the projection of the Weyl curvature of the black
hole on the brane. This source can be interpreted as a CFT in a thermal
state corresponding to the Hawking temperature of the bulk black hole.
The brane cosmological metric has a constant curvature spatial part,
and its symmetries demand that only a radiation energy Weyl term is
allowed. From the bulk perspective, this means that every point on
the brane is at the same distance from the bulk black hole. Thus a 
flat universe corresponds to a `flat' bulk black hole, a closed 
universe to a conventional spherical bulk black hole. One way of 
imagining a braneworld black hole forming is to transport this 
bulk black hole in towards and onto the brane. In doing this, one
breaks this symmetry of equidistance from the black hole. In other
words, from the holographically dual brane point of view, we introduce
an anisotropic stress by shifting the black hole from its equidistant
position. The closer we bring the black hole towards the brane, the
more anisotropic the set-up. Therefore, we might expect that $\Pi$ 
becomes more and more important both as we transport the black hole
towards the brane, or as we move closer to the event horizon. 
Therefore a physically reasonable expectation might be that equations
of state with large $\gamma$ are relevant near the horizon.

Taking this reasoning to its logical extreme, we therefore propose 
as a ``working metric'' for the near horizon solution the analytic
form of section \ref{sub:ansol}:
\be
ds^2 = {(r-r_1)^2\over (r+r_0)^2} dt^2 - {(r+r_0)^4\over r^4} dr^2
- {(r+r_0)^4\over r^2} d\Omega_{I\!I}^2 \label{wormmet2}
\ee

Clearly, although there is a degree of arbitrariness in this choice, 
we believe that the horizon is likely to be singular, and also that
a turning point in the area function is also likely. This metric
has both these features and the added advantage of analyticity, which
means that many properties can be calculated explicitly.

\section{Testing the solution and future horizons}\label{sec:bhsol}

We now wish to explore some simple tests of the near horizon form
(\ref{wormmet}/\ref{wormmet2}) or (\ref{areaworm}):
\be
ds^2 = \left [ (1+\epsilon) \sqrt{1-{2GM\over R}} - \epsilon \right ]^2
dt^2 - {dR^2 \over 1-{2GM\over R}} - R^2 d\Omega_{I\!I}^2
\label{areaworm2}
\ee

Although this latter form of the metric is not valid throughout the
exterior of the horizon if $\epsilon <0$, it is an easier gauge to
see the contrasts with the Schwarzschild metric, and since it turns
out that there can be no stable orbit within the wormhole region, it
is quite satisfactory for working with accretion discs.

The first point to note about (\ref{areaworm2}) is that the
ADM and gravitational mass (defined by $g_{tt}$) are no longer
the same. Once again, this is quite a common occurance in string
gravity, and is a result of the extra degrees of freedom of the
gravitational field. As a result, the weak field tests of light
bending and perihelion precession will be modified at the 
$O(\epsilon)$ level, in a way consistent with the PPN formalism.

The near horizon metric (2.24 in area gauge or 4.4 in isotropic
coordinates) is very different to the standard Schwarzschild
solution. The horizon is {\em always} singular even for apparently
negligibly small $\epsilon$ (except for $\epsilon=0$ which corresponds
to $\cal E{\mu\nu}=0$, so is identically Schwarzschild). This is a true
singularity, as the energy density from $\cal E_{\mu\nu}$ becomes
infinite at the horizon. An intrepid observer plunging into a
supermassive black hole, expecting to sail seamlessly through the
horizon and explore spacetime close to the singularity before finally
succumbing to tidal forces, is instead crushed out of existence {\em
at} the horizon. Indeed, they are slammed into the horizon at infinite
proper speed! The are no timelike or null geodesics which connect
anywhere outside the horizon with the standard Schwarzschild
singularity at $r=0$.  Infalling matter or light simply cannot reach
this point.

Such a drastic change to the spacetime surely has observable effects
{\em above} the horizon. The metric (4.4) can easily be compared with
the PPN formalism, and observational limits on these parameters from
solar system tests set limits on $|\epsilon|< 10^{-3}$. However,
these derive from solar system tests in the weak gravitational field
regime. This would only be applicable if the metric (4.4) covered
the entire spacetime, but as discussed above we envisage this only as
the near horizon asympote of a more general metric. The only mathematical
limit on the near horizon metric is $\epsilon>-\frac{1}{2}$ from
the requirement that $r_1$ is positive definite.

Here instead we constrain $\epsilon$ by connection to the observations
of accreting black holes in our galaxy. The luminosity and temperature
of the blackbody emission from the accretion disc can be measured from
X-ray spectra, giving a direct estimate of the emitting area. 
There are uncertainties in this approach, but some confidence can be 
derived from the fact that changes in the luminosity give rise to 
the $L\propto T^4$ behaviour expected from a constant emitting area. 
Assuming that the uncertainties are less than a factor of 2 then the
data strongly constrain the minimum stable orbit to be $<12GM$.

We assume that the `near horizon' metric applies on these size scales
and solve the Euler-Lagrange equations from (2.24) in the equatorial
plane to find the innermost stable particle orbit, and its associated
angular momentum $h=|g_{\phi\phi}|\dot{\phi}$. Table 1 shows these as a
function of $\epsilon$, together with the horizon position
($g_{tt}=0$) and the radius at which light can (unstably) orbit around
the black hole. We can use the area gauge even for cases where
$\epsilon<0$ (which have a wormhole) as the coordinate singularity at
$R=2GM$ is below the radii of interest for light and particle orbits.
The observations of accretion disc size in X-ray binary systems
give an upper limit to $\epsilon$ of about $2$, and are easily
consistent with our expectation of $\epsilon<0$. (In the absence of
a complete metric from larger to smaller $r$, we have taken $M$ to
be the ADM mass.)

\begin{table*}
\begin{tabular}{lccccccc}
\hline
$\epsilon$ & -0.5 & -0.1 & 0 & 0.1 & 0.5 & 1 & 2 \\
\hline
$R_h$ & $\infty^*$ & 2.02$^*$ & 2 & 2.01 & 2.25 & 2.67 & 3.6 \\
$R_{ms}$ & 4.26 & 5.63 & 6 & 6.37 & 7.88 & 9.82 & 13.75\\
$h^2$ & 4.15 & 10.11 & 12 & 14.03 & 23.78 & 39.54 & 83.06\\
$R_{ph}$ & 2.25 & 2.83 & 3 & 3.17 & 3.91 & 4.86 & 6.82\\

\hline
\end{tabular}

\caption{Radii for the horizon, $R_h$, 
minimum stable orbit for particles, $R_{ms}$ and the unstable photon
orbit $R_{ph}$ in units of $GM$. We 
also give the associated angular momentum of the 
minimum stable orbit, $h^2$. Note that for negative $\epsilon$, the
star indicates that the horizon is on the lower Kruskal branch, and
therefore {\it inside} R=2GM.}
\end{table*}

If this asymptotic metric holds to within a few tens of Schwarzschild
radii of the horizon then there are potentially observable effects on
the spacetime around stellar mass black holes. We will explore these
further in a subsequent paper.

Although we have focused on braneworlds, the techniques will clearly
be applicable to stringy black holes, or indeed any theory that
modifies gravity. In the absence of a concrete solution to test, our
approach of trying to find classes of behaviour and some sort
of universality near the event horizon seems the best way to proceed.
Typically, any covariant theory which modifies the near horizon 
spacetime structure of the black hole will have these qualitative
features, hence similar effects that we have been exploring. All in all,
this seems a fruitful alternative to explore in terms of testing ideas
in high energy gravity.

\acknowledgments
We would like to thank Roy Maartens for useful conversations.

\end{document}